\documentclass[a4paper,12pt, epsfig]{article}

\usepackage{epsfig}
\usepackage{amssymb}
\usepackage{amsfonts}
\usepackage{amsmath}

\newskip\humongous \humongous=0pt plus 1000pt minus 1000pt

\newif\ifdtup

\catcode`\@=11

\@addtoreset{equation}{section}
\def\theequation{\thesection.\arabic{equation}}

\def\@normalsize{\@setsize\normalsize{15pt}\xiipt\@xiipt
\abovedisplayskip 14pt plus3pt minus3pt%
\belowdisplayskip \abovedisplayskip
\abovedisplayshortskip \z@ plus3pt%
\belowdisplayshortskip 7pt plus3.5pt minus0pt}

\def\small{\@setsize\small{13.6pt}\xipt\@xipt
\abovedisplayskip 13pt plus3pt minus3pt%
\belowdisplayskip \abovedisplayskip
\abovedisplayshortskip \z@ plus3pt%
\belowdisplayshortskip 7pt plus3.5pt minus0pt
\def\@listi{\parsep 4.5pt plus 2pt minus 1pt
     \itemsep \parsep
     \topsep 9pt plus 3pt minus 3pt}}

\relax

\catcode`@=12

\catcode`\@=11

\def\section{\@startsection{section}{1}{\z@}{3.5ex plus 1ex minus
   .2ex}{2.3ex plus .2ex}{\large\bf}}

\def\thesection{\arabic{section}}
\def\thesubsection{\arabic{section}.\arabic{subsection}}

\def\appendix{\setcounter{section}{0}
 \def\thesection{Appendix \Alph{section}}
 \def\thesubsection{\Alph{section}.\arabic{subsection}}
 \def\theequation{\Alph{section}.\arabic{equation}}}
\def\SymBoxes#1#2#3#4{\newdimen\un@t \un@t#3%
\raisebox{#1}{\rule{#2\un@t}{#4}\hskip-#2\un@t% lower horizontal
\@tempdimb\un@t \advance\@tempdimb by-#4\@tempcntb#2\relax%
\@whilenum{\@tempcntb>0}\do{%                         % #2 vertical lines
\rule{#4}{\un@t}\hskip\@tempdimb \advance\@tempcntb by\m@ne}%
\hskip-#2\un@t \rule[\un@t]{#2\un@t}{#4}%
\rule[\un@t]{#4}{#4}\hskip-#4%             % upper horizontal line
\rule{#4}{\un@t}}\hskip-#4}                % rightest vertical line
\begin{document}
%\begin{letter}{~}

%%%%%%Define some new commands and  macros

\newcommand{\beq}{\begin{equation}}
\newcommand{\eeq}{\end{equation}}

\newcommand{\bea}{\begin{eqnarray}}
\newcommand{\eea}{\end{eqnarray}}

\newcommand{\beas}{\begin{eqnarray*}}
\newcommand{\eeas}{\end{eqnarray*}}
\newcommand{\non}{\nonumber}
\newcommand{\bquo}{\begin{quote}}
\newcommand{\enqu}{\end{quote}}

%%%%%%%%%%%%%%%%
\renewcommand{\(}{\begin{equation}}
\renewcommand{\)}{\end{equation}}

%%%%%%definitions

\def\Nc{N_\textrm{c}}
\def\Nf{N_\textrm{f}}

\def\n{\mathbf{n}}
\def\m{\mathbf{m}}

\def\half{\frac{1}{2}}

\def\IZ{{\mathbb Z}}
\def\IR{{\mathbb R}}
\def\IC{{\mathbb C}}
\def\IQ{{\mathbb Q}}

\def \eqn#1#2{\begin{equation}#2\label{#1}\end{equation}}

\def \d{\textrm{d}}
\def \p{\partial}
\def\Tr{ \hbox{\rm Tr}}

\def\const{\hbox {\rm const.}}

\def\bra{\langle}\def\ket{\rangle}
\def\Arg{\hbox {\rm Arg}}

\def\diag{\hbox{\rm diag}}

\def\a{\alpha}

\def\tq{{\widetilde q}}
\def\pr{\prime}

\def\hsp{,\hspace{.7cm}}

\begin{titlepage}

\begin{flushright}
ULB-TH/08-23\\
TAUP 2879-08\\
%hep-th/yymmnnn\\
\end{flushright}

\bigskip

\def\thefootnote{\fnsymbol{footnote}}

\begin{center}
{\Large {\bf A New Holographic Model of Chiral Symmetry Breaking}} \\
\end{center}

\bigskip

\begin{center}

{\large Stanislav Kuperstein\footnote{skuperst@ulb.ac.be}}

\end{center}

\begin{center}
\textit{{\it Physique Th\'eorique et Math\'ematique,
International Solvay
Institutes, \\ Universit\'e Libre de Bruxelles, ULB Campus Plaine C.P.
231, B--1050 Brussel, Belgi\"e}}
\end{center}

\begin{center}

{\large Jacob Sonnenschein\footnote{ cobi@post.tau.ac.il}}

\end{center}

\begin{center}
\textit{{\it School of Physics and Astronomy,\\
The Raymond and Beverly Sackler Faculty of Exact Sciences,\\
Tel Aviv University, Ramat Aviv, 69978, Israel.}}
\end{center}

\vfil

\renewcommand{\thefootnote}{\arabic{footnote}}

\noindent
\begin{center} {\bf Abstract} \end{center}

A new  family of models of flavour chiral symmetry breaking is
proposed. The models are  based on the embedding of a stack of $D7$
branes  and a stack of anti-$D7$ branes in the conifold background.
This family of  gravity models is dual to a field theory with
spontaneous breaking of conformal invariance and chiral
flavour symmetry. We identify the corresponding Goldstone bosons and
compute the spectra of massive scalar and vector mesons. 
The dual quiver gauge theory is also discussed. We further analyse a model where
chiral symmetry is not broken.

\vfill

\begin{flushleft}
{\today}
%\vspace{1cm}
\end{flushleft}

\end{titlepage}
%\bigskip

\vspace{1.6 cm}

\vfill

\bigskip

\hfill{}
\bigskip

\tableofcontents

\setcounter{footnote}{0}

%%%%%%%%%%%%%%%%%%%%%%%%%%%%%%%%%%%%%%%%%%%%%%%%%%%%%%%%%%%%%

\section{\bf Introduction and Summary}

%\noindent

Though the recipe for building the string theory of QCD and hadrons
is still a mystery, it should certainly include the ingredients of
confinement and flavour chiral symmetry and its spontaneous
breakdown. Whereas the realization of the former is easy, the
incorporation of the latter is not and  is shared  only by  very few
models.

Holographic models are based on taking the near horizon limit of the
background produced by  large $\Nc$ branes. Adding  $\Nf$ additional
branes, introduces   strings  stretching between the two type of
branes that transform in the fundamental representation of
$U(\Nc)\times U_L(\Nf)$. Thus, for $\Nf \ll \Nc$, when the
back-reaction of the additional branes on the background can be
neglected,
 placing a
stack of  $\Nf$ $D$-branes in a holographic background associates
with  adding  fundamental quarks in the dual gauge theory.  Putting
now an additional stack of $\Nf$ \emph{anti} $D$-branes  results in
\emph{anti}-quarks that transform in the fundamental representation
of another $U_R(\Nf)$ symmetry which is a gauge symmetry  on the new
stack of branes. In such a setup the dual gauge theory enjoys the
full $U(\Nf)_L \times U(\Nf)_R$  flavour symmetry. However,
if the branes and the anti-branes \emph{smoothly} merge at some
point into a single configuration then only a single
$U(N)_\textrm{D}$ factor survives. If one can attribute the region
where the two separate symmetry groups reside  to the UV regime  of
the dual field theory and where they merge to the IR , then one
achieves a ``geometrical mechanism" in the gravity model  dual of
the gauge theory  chiral symmetry breakdown.

Such a scenario was derived by adding $D7$ and anti- $D7$ branes to
the confining Klebanov-Strassler background (KS)
\cite{Klebanov:2000hb} model in \cite{Sakai:2003wu}. Holomorphic
embeddings of $D7$ branes into the KS model  which is dual to
supersymmetric gauge theory without flavour chiral symmetry breaking
were studied in \cite{Ouyang,Levi:2005hh} and
\cite{Me}.
The backreaction of the flavour branes (in the so-called un-quenched approximation)
have been further investigated in a series of papers
\cite{Benini:2006hh,Benini:2007gx,Benini:2007kg,Bigazzi:2008zt,Chen:2008jj}.

A similar geometrical mechanism was implemented in the
Sakai-Sugimoto model \cite{Sakai:2004cn,Sakai:2005yt}. This model
incorporates  $\Nf$ $D8$ and $\bar{D}8$ probe branes into Witten's
model \cite{Witten:1998zw} which is based on the near extremal
$D4$-brane background. An analogous non-critical six dimensional
flavoured model was written in \cite{Casero:2005se} and \cite{Mintkevich:2008mm} using the
construction of \cite{Kuperstein:2004yk,Kuperstein:2004yf}.

Despite its tremendous success the Sakai-Sugimoto model
\cite{Sakai:2004cn}  suffers from various drawbacks which it
inherits from Witten's model \cite{Witten:1998zw}. In particular the
model is inconsistent in the UV region due to the fact that the
string coupling diverges there. In addition the dual field theory is
in fact a five dimensional gauge theory compactified on a circle
rather than  a four dimensional gauge theory. A potential way to
bypass these problems is to use as a background the KS model since
it is based on $D3$ branes and its dilaton does not run. As
mentioned above this was the main idea behind \cite{Sakai:2003wu}.
However, the solution found there for the classical probe profile
included an undesired gauge field on the transverse $S^3$.  On the
route to deriving novel solutions of  the embedding of $D7$ and
anti-$D7$ branes in the KS model, the goal of the present paper is
to solve for the embedding of these flavour branes in the context of
the un-deformed conifold geometry. The $10d$ solution based on this
geometry is known as the Klebanov-Witten (KW) background
\cite{{Klebanov:1998hh}}.

The summary of the achievements of the paper are the following:
\begin{itemize}
\item
We write down the DBI action associated with the embedding of $D7$
branes  in the geometry of $AdS_5\times T_{11}$. We write the
corresponding equations of motion associated with the two angles on
the $S^2$ which is  transverse to  the probe branes. We find an
analytic solution  for the classical embedding. In fact it is a
family of profiles  along the equator of the $S^2$ which are
characterised by the minimal radial extension of the probe brane
$r_0$ and with an asymptotic fixed span of $\sqrt{6}\pi/4$ for the
equatorial angle.
\item
We introduce a Cartesian-like coordinates that enable us to examine
the spectrum of  scalar mesons associated with the fluctuations of
the embedding.
\item
We identify a massless mode that plays the role of the Goldstone
boson associated with the spontaneous breakdown of conformal
invariance.
\item
We compute the spectrum of the massive  vector mesons.
\item
We identify the ``pions" associated with the chiral symmetry
breaking. They are the zero modes of the gauge fields along the
radial direction.
\item
We write down the quiver that describes the dual gauge field.
We also argue why our model includes Weyl and not Dirac fermions as required for a model 
with chiral symmetry breaking.
\item
We describe a special case where chiral symmetry is not broken.
\end{itemize}

The paper is organised as follows: In Section \ref{Thesetup} we present the basic
setup of the model. Section \ref{Theconfiguration} is devoted to the derivation of the $D7$
probe brane profile solution. We start with a brief review of the
conifold geometry. We then write the DBI action and solve the
corresponding equation of motion. The spectrum of mesons is
extracted in Section \ref{Spectrum}. We identify the Goldstone mode associated
with the spontaneous breaking of conformal invariance and the
``pions" that follow from the breaking of the flavour chiral
symmetry. We further derive the spectrum of massive vector and
scalar mesons. Section \ref{GT} is devoted to the dual field theory. We
draw the corresponding quiver diagram and discuss the properties of
the theory. In Section \ref{z4} we discuss a special model where chiral
symmetry is not broken.

%%%%%%%%%%%%%%%%%%%%%%%%%%%%%%%%%%%%%%%%%%%%%%%%%%%%%%%%%%%%%

\section{The basic setup}
\label{Thesetup}

To understand the basic setup of the $D7$-branes in the conifold
geometry we first  review the setup of  the type IIA  model of
\cite{Sakai:2004cn}. As was mentioned above it is based on adding to
Witten's model \cite{Witten:1998zw} a stack of $\Nf$ $D8$ branes and
a stack of $\Nf$ anti-$D8$ branes. The $D8$-branes are $9d$ objects,
which means that there is only one coordinate transversal to them.
Asymptotically this coordinate  $x_4$  is actually one of
world-volume coordinates of the original $D4$ branes.  The
coordinate is along an $S^1$  compactified direction. The
submanifold of the background along this direction and the radial
direction has a ``cigar-like" shape. The radius of the cycle shrinks
to zero size at some value of the radial direction  $u=u_\Lambda$
and diverges asymptotically for large $u$. The profile of the $D8$
probe branes, which is determined by the equations of motion deduced
from the DBI action, is of a $U$-shape. It stretches from $x_4=-L/2$
at $u \rightarrow \infty$ down to $x_4=0$ at a minimum value of
$u=u_0 \geqslant u_\Lambda$ and back to $x_4=+L/2$ at asymptotic
$u$. This shape is obviously in accordance with the fact that on the
``cigar" geometry there is no way for the $D8$ branes and the anti
$D8$ to end. Their only choice is to merge. Slicing the cigar at
large $u$ we have two distinct branches of $D8$ branes with
$U(\Nf)_L$ gauge field of the left one and $U_R(\Nf)$ on the right
one. This is the dual picture of the full chiral symmetry at the UV
region of the gauge theory. On the other hand down at the tip of the
$U$-shape there is only a single $U(\Nf)_\textrm{D}$ gauge symmetry
which stands for the unbroken global symmetry in the dual gauge
theory. Thus the gravity dual of the spontaneous breakdown of chiral
symmetry is the $U$-shape structure of the probe branes. A given
probe brane profile is characterised by $L\sim 1/\sqrt{u_0}$. We
mention this relation to contrast the situation that will be found
for  the $D7$ branes on the conifold. In terms of the dual gauge
theory the separation distance $L$ is related to the mass of the
mesons. For configurations with $u_0 \gg u_\Lambda$ one finds that
the meson mass behaves like $1/L$. Flavour chiral symmetry
restoration occurs in QCD at high temperature at the deconfining
phase of the theory. In the dual gravity model
\cite{Aharony:2006da, Parnachev:2006dn,Peeters:2006iu} this phase is described
by a distinct geometry of the background where the cigar-like shape
describes the submanifold of the Euclidean time direction and the
radial direction whereas the $(x_4,u)$ slice has now a shape of a
cylinder that stretches from some minimal value $u=u_\textrm{T}$ to
infinity. In this geometry the two separate stacks of branes have
two options: either to merge like in the low temperature phase or to
reach an end separately. The former case translates into a
deconfining phase which chiral symmetry breakdown and the latter
corresponds to a deconfined phase with a restoration of the full
flavour chiral symmetry.

\begin{figure}[!h]
\begin{center}
\includegraphics[scale=0.5
]{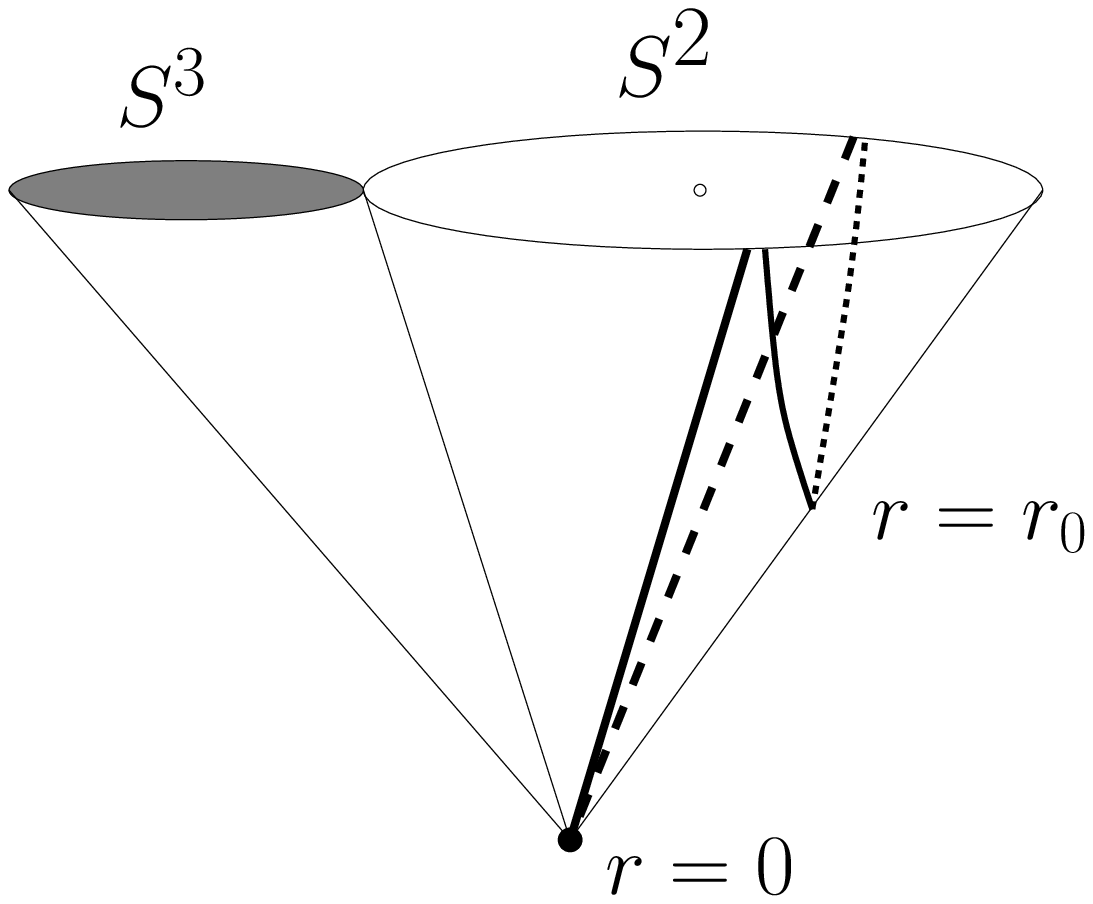}
\caption{The picture shows two possible $D7$ configurations. In both cases the branes wrap the $S^3$ and
look like two separate points on the $S^2$. The position of this points depends, however, on the radial coordinate $r$.
In one case (the $V$-shape) the stacks meet only at the tip of the conifold (the thick curve), while in the other
(the $U$-shape) they merge already at
$r_0 > 0$ (the thin curve).}
\label{Profile}
\end{center}
\end{figure}

Since we deal with the type IIB supergravity we will need instead a
pair of $D7$-branes. Now the transversal space is two-dimensional
and analogously to the Sakai-Sugimoto model we need a two-sphere to
place the branes on. This is indeed the case as the $T^{1,1}$ base of the
conifold has an $S^3 \times S^2$ topology. We now have two different
options for the $D7$-brane configuration. One possibility is to
place the branes at two separate points on the two-sphere and
stretch them to the tip of the conifold, where the two-sphere and
the three-sphere shrink. We will refer to this configuration as a
$V$-shape. Another possibility is a $U$-shape configuration with
$D7$-branes smoothly merging into a single stack at some point
$r=r_0$  along the radial direction away from the tip. The two
options are depicted on Figure \ref{Profile}.

\begin{figure}[!h]
\begin{center}
\includegraphics[scale=0.6
]{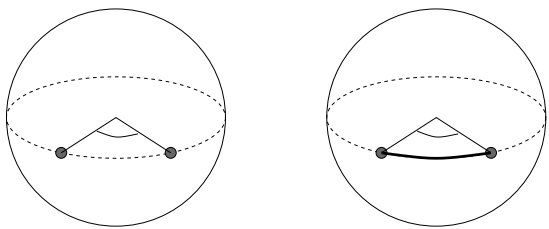}
\caption{The picture shows two different $D7$-brane profiles on the two-sphere.
For the $V$-shape configuration  (left) the $D7$-branes are given by two separate \emph{fixed} points on the $S^2$, while for
the $U$-shape (right) the position of the two points along the equator depends on $r$ and they are smoothly connected
in the middle of the arc for $r=r_0$.}
\label{S2}
\end{center}
\end{figure}

We claim that the configuration reaching the tip describes the
chiral symmetric phase, while the $U$-shape configuration ending at
$r_0$ corresponds to the broken chiral symmetry. It looks somewhat
perplexing, since instead of a pair of two \emph{parallel}
$D7$-branes we have $D7$-branes that still meet at the tip. Notice,
however, that the tip is necessarily a singular point and so the two
branches of the $V$-shape are ``distinguishable"  and correspond to
two separate branes. Putting it more bluntly, the tip is a
co-dimension six point (both the $S^2$ and the $S^3$ shrink there!),
so the right way to analyse the configuration is to consider its
form in the full $10d$ background. The radial coordinate of the
conifold combines then with the space-time coordinates to build
$AdS_5$, which is completely wrapped by the $D7$-branes. The branes
wrap also the three-sphere. On the two-sphere, on the other hand,
for the $V$-shape the branes look like two separate fixed point,
while the $U$-shape corresponds to an arc along the equator. The
situation is shown on Figure \ref{S2}.

An important issue related to the position of the brane on the
two-sphere is the amount of supersymmetry preserved by the probe
branes. One might think that the two stacks should be located at the
antipodal points, let's say the north and the south pole. In such a
case the embedding is holomorphic (see Section \ref{z4}) and so the
setup preserves some supersymmetry. 
This na\"ive expectation, however, proves to be wrong,
since the configuration with two antipodal points does not solve the
equations of motion as we will see in the next section.

%%%%%%%%%%%%%%%%%%%%%%%%%%%%%%%%%%%%%%%%%%%%%%%%%%%%%%%%%%%%%

\section{The configuration}
\label{Theconfiguration}

In this section we solve the equations of motion for the $D7$-brane deriving the $U$-shape discussed above.
The solution involves a free parameter $r_0$ which is just the minimal value of radial coordinate along the profile.
As $r_0$ goes to zero we will find the $V$-shape configuration. The latter, as we have explained earlier,
corresponds actually to a pair of two separate $D7$-branes.
We start our journey by reviewing the conifold basics (for a more detailed explanation see
\cite{Evslin:2007ux}).

\subsection{\bf Brief review of the conifold geometry}

The conifold is a $3d$ complex subspace inside $\mathbb{C}^4$ defined by a $2 \times 2$ matrix $W$
with vanishing determinant ($\det W=0$). Since the definition is obviously scaling invariant
we can fix the radial coordinate of the conifold as:
\beq
\label{rho:eq}
\rho^2 = \textrm{Tr} \left( W^\dagger W \right).
\eeq
Here $\rho$ and the more common radial coordinate $r$ are related by:
\beq
\label{eq:rho-r}
\rho^2 = \frac{2^{5/2}}{3^{3/2}} r^3.
\eeq
Because $W$ is singular it necessarily has one left and one right null eigenvectors. This in turn implies that $W$
can be re-cast in the form:
\beq
\label{Wuv:eq}
W =\rho u v^\dagger,
\eeq
where the vectors $u$ and $v$ both have length one ($u^\dagger u=v^\dagger v=1$).
With these notations the null eigenvectors are $u^{\textrm{T}}\epsilon$ and $\epsilon v^\star$, where $\epsilon$
is the $2 \times 2$ anti-symmetric tensor.  The representation (\ref{Wuv:eq}) is of course not unique,
since $W$ is invariant under:
\beq
\label{uvuv:eq}
u \to e^{i \varphi} u, \qquad  v \to e^{i \varphi} v.
\eeq
This way we arrive at a different, but equivalent, definition of the conifold. It can be defined
as a \emph{K\"ahler quotient} of
$\mathbb{C}^4$ with the $U(1)_K$ gauge charges $(1,1,-1,-1)$. Denoting the $\mathbb{C}^4$ coordinates by ${\mathfrak{z}}_1$,
${\mathfrak{z}}_2$, ${\mathfrak{z}}_3$ and ${\mathfrak{z}}_4$ we easily find  that
$u=\sqrt{\rho} ({\mathfrak{z}}_1,{\mathfrak{z}}_2)^{\textrm{T}}$ and
$v=\sqrt{\rho} ({\bar{\mathfrak{z}}}_3,{\bar{\mathfrak{z}}}_4)^{\textrm{T}}$. Let us now introduce a $2 \times 2$
matrix $X$ satisfying:
\beq
\label{uXv:eq}
u = Xv.
\eeq
If we also impose an additional constraint saying that $X$ is special and unitary (namely $X \in SU(2)$), then there is
an \emph{unique} solution for (\ref{uXv:eq}), given by $X = u v^\dagger - \epsilon u^\star v \epsilon$.
Since $X$ is clearly invariant under (\ref{uvuv:eq}) we see that $X$ parameterizes an $S^3$. Furthermore,
using the Hopf map
we realize that the $U(1)_K$ transformation (\ref{uvuv:eq}) implies that the unit length vector $v$ alone
defines an $S^2$. Starting with $X$ and $v$ we can find $u$ and then $W$. We get:
\beq
\label{WXvv:eq}
W = \rho X v v^\dagger.
\eeq
We conclude that $T^{1,1}$, 
the base of the conifold (the slice given by $\rho=\textrm{const}$), is uniquely
parameterized by $X$ and $v$, so the topology of the base is indeed $S^3 \times S^2$.

Let us now make contact with the explicit $S^3 \times S^2$ conifold
coordinates used in the literature
\cite{Minasian:1999tt,Gimon:2002nr,Krishnan:2008gx}. First, note
that $v v^\dagger$ is a hermitian  matrix with eigenvalues $1$ and
$0$. We therefore can write: \beq v v^\dagger = \mathcal{V} \left(
\begin{array}{cc} 1 & 0 \\ 0 & 0 \end{array}
\right)\mathcal{V}^\dagger, \eeq where $\mathcal{V}$ is an $SU(2)$
matrix fixed by $v$ up to the gauge transformation $\mathcal{V} \to
\mathcal{V} e^{i \varphi \sigma_3}$. Exactly like for $v$ the matrix
$\mathcal{V}$ defines an $S^2$ by virtue of the Hopf map. Second, we
set: \beq \mathcal{V}=e^{\frac{i}{2} \phi \sigma_3 } e^{\frac{i}{2}
\theta \sigma_2 }. \eeq It is always possible to bring the matrix
$\mathcal{V}$ to this form using a gauge transformation. We are
finally in a position to write the conifold metric in the $S^3
\times S^2$ coordinates: 
\beq \label{metric6:eq} 
\d s_{(6)}^2 = \d r^2
+ \frac{r^2}{3} \left( \frac{1}{4} (f_1^2+f_2^2) + \frac{1}{3} f_3^2
+
 ( \d \theta - \half f_2)^2 + ( \sin \theta \d \phi - \half f_1)^2 \right),
\eeq
where $r$ was introduced in (\ref{eq:rho-r}) and the $1$-forms $f_i$ are defined as:
\beq
\label{eq:hw}
\left( \begin{array}{c} f_1 \\ f_2 \\ f_3 \end{array} \right) =
\left( \begin{array}{ccc} 0 & \cos \theta & -\sin \theta \\ 1 & 0 & 0 \\
0 & \sin \theta & \cos \theta \end{array} \right)
\left( \begin{array}{ccc} -\sin \phi & -\cos \phi & 0 \\ -\cos \phi & \sin
\phi & 0 \\ 0 & 0 & 1\end{array} \right)
\left( \begin{array}{c} w_1^\prime \\ w_2^\prime \\ w_3^\prime \end{array} \right),
\eeq
where $w_i$'s are the $SU(2)$ left-invariant Maurer-Cartan
one forms\footnote
{The $S^3$ matrix $T$ and the $S^2$ matrix $S$ of \cite{Minasian:1999tt,Gimon:2002nr} are related to $X$
and $\mathcal{V}$ through $T=X \sigma_3$ and
$S=\sigma_3 \mathcal{V} \sigma_3=e^{\frac{i}{2} \phi \sigma_3 } e^{-\frac{i}{2} \theta \sigma_2}$. The Maurer-Cartan
forms determined by $T^\dagger \d T = \frac{i}{2} \sigma_i w_i$ are related to $w_i^\prime$'s as follows:
$w_{1,2}=-w_{1,2}^\prime$ and $w_{3}=w_{3}^\prime$.}:
\beq
X^\dagger \d X = \frac{i}{2} \sigma_i w_i^\prime.
\eeq
The two $SO(3)$ matrices in (\ref{eq:hw}) reflect the fact that the
three-sphere is fibered over the two-sphere. This fiber is
trivial as one can easily verify by properly calculating the Chern class
of the fiber bundle\footnote{For what follows it will be useful to note that
$\sum_{i=1}^3 f_i^2=\sum_{i=1}^3 {w^\prime}_i^2$
and $f_1 \wedge f_2 \wedge f_3 = w_1^\prime \wedge w_2^\prime \wedge w_3^\prime$.}.

Let us end this section with a remark on the un-deformed conifold symmetries.

First, there is a $\IZ_2$ symmetry that acts as $W \to
W^\textrm{T}$. On the gauge theory side the symmetry replaces the
two $SU(\Nc)$ gauge groups. This fact becomes obvious if following
\cite{Klebanov:1998hh} one identifies the K\"ahler quotient
coordinates $\mathfrak{z}_i$ with the bi-fundamental chiral
superfields $A_{1,2}$ and $B_{1,2}$: \beq \label{zzAB:eq}
(\mathfrak{z}_1,\mathfrak{z}_2) = \left( A_1,A_2 \right) \qquad
\textrm{and} \qquad (\mathfrak{z}_3,\mathfrak{z}_4) = \left( B_1,B_2
\right). \eeq Since under $W \to W^\textrm{T}$ we have
$(\mathfrak{z}_1,\mathfrak{z}_2) \leftrightarrow
(\mathfrak{z}_3,\mathfrak{z}_4)$, the fields $A_i$ and $B_i$ are
also interchanged. These fields transform in the $(\mathbf{\Nc},
\bar{\mathbf{\Nc}})$ and $(\bar{\mathbf{\Nc}},\mathbf{\Nc})$
representations of the $SU(\Nc) \times SU(\Nc)$ gauge group, and so
the $\IZ_2$ interchanges also the $SU(\Nc)$'s. On the other hand,
from (\ref{Wuv:eq}) and (\ref{uXv:eq}) we have $(u,v) \to
(v^\star,u^\star)$ or alternatively $(X,v) \to (X^\textrm{T}, (X
v)^\star)$ under $\IZ_2$. This means that our configuration (Figure
\ref{Profile}) which will be discussed in details below, 
certainly breaks the $\IZ_2$ symmetry. 
It follows from the fact that $v$ parameterizes the $2$-sphere
and the position of the brane on the $S^2$ depends only on the radial coordinate
and not on $X$, and so the $\IZ_2$ transformation of $v$ is not respected by out setup.
This conclusion will play
an important r\^ole in the gauge theory discussion in Section
\ref{GT}.

Second, there is an $SU(2)_1 \times SU(2)_2$ symmetry that acts as $W \to S_1 W S_2^\dagger$, where
$S_1$ and $S_2$ are two $SU(2)$ matrices. Under this symmetry the fields $A_i$ and $B_i$ transform as
a doublet of one $SU(2)_i$ factor and as a singlet of the other. From (\ref{WXvv:eq}) we see that
$(X,v) \to (S_1 X S_2^\dagger, S_2 v)$ and so our embedding breaks $S_2$, but not $S_1$.
This fact is expected, since the broken $\IZ_2$ from the previous paragraph interchanges the two $SU(2)_i$
symmetries. If, for instance, we were using $u$ (and not $v$) to parameterize the two-sphere, then $S_1$
would be broken instead (and not $S_2$).

\subsection{\bf The $D7$ brane profile}

In this paper we will study a $D7$-brane configuration, which spans
the space-time coordinates $x_\mu$, the radial direction $r$ and the
three-sphere parameterized by the forms $f_i$ (or alternatively
$w_i$). The transversal space is given by the two-sphere coordinates
$\theta$ and $\phi$. Remarkably, since $w_i$ are left-invariant
forms, our \emph{ansatz} preserves one of the $SU(2)$ factors of the
global symmetry of the conifold. Based upon this observation, we
will assume that  $\theta$ and $\phi$ do not depend on the $S^3$
coordinates. Since our profile still breaks one $SU(2)$ this
assumption should  be examined more carefully. Upon expanding the
action around the solution we will find  that the contributions of
the non-trivial $S^3$ modes appear only at the second order at the
fluctuations\footnote{ Notice that in doing so we have also to
include the contributions coming from the variations of the $SO(3)$
matrices in (\ref{eq:hw}). This, however, does not modify the final
conclusion.}. We, therefore, can safely assume that along the
classical profile $\theta$ and $\phi$ depend only on the radial
coordinate.

The $10d$ metric is: \beq \label{metric10d:eq} \d s_{(10)}^2 =
\frac{r^2}{R^2} \d x_\mu \d x^\mu + \frac{R^2}{r^2} \d s_{(6)}^2
\eeq with the $6d$ metric given by (\ref{metric6:eq}) and the
$AdS_5$ radius is $R^4=\frac{27}{4} \pi g_s \Nc \ell_s^4$. Because the KW
background has no fluxes except for the $C_4$ form the Chern-Simons
terms do not contribute and the action consists only of the DBI
part: \beq \label{DBI:eq} S_{\textrm{DBI}}=-\mu_7 \int \sqrt{-g_8}.
\eeq Substituting $\theta=\theta(r)$ and $\phi=\phi(r)$ into the
metric we find the following Lagrangian: \beq \mathcal{L} \propto
r^3 \left( 1 + \frac{r^2}{6} \left( \theta_r^2 + \sin^2 \theta
\phi_r^2 \right) \right)^{1/2}. \eeq Here the subscript $_r$ stands
for the derivatives with respect to $r$. The Lagrangian is $SU(2)$
invariant, so we can restrict the motion to the equator of the
two-sphere parameterized by $\theta$ and $\phi$. Setting
$\theta=\pi/2$ we easily find the solution of the equation of
motion\footnote {There are two initial parameters we have to fix in
the solution: one is $r_0$ and the other is the value of $\phi$ at
$r=r_0$, which we set to $0$.}: \beq \label{Solution:eq} \cos \left(
\frac{4}{\sqrt{6}} \phi(r)\right) = \left( \frac{r_0}{r} \right)^4.
\eeq There are two branches of solutions for $\phi$ in
(\ref{Solution:eq}) with $\phi \in [-\pi/2,0]$ or $\phi \in
[0,\pi/2]$. For $r_0=0$ we have two fixed ($r$-independent)
solutions at $\phi_-=-\frac{\sqrt{6}}{8} \pi$ and
$\phi_+=\frac{\sqrt{6}}{8} \pi$. The induced $8d$ metric in this
case is that of $AdS_5 \times S^3$ as one can verify\footnote
{\label{squashed}
To be more precise the transversal space is $S^3$ only topologically since
not all the coefficients of $f_i^2$'s in (\ref{metric6:eq}) are
equal. This is rather a ``squashed" $3$-sphere.} by plugging $\d
\phi=\d \theta=0$ into (\ref{metric6:eq}). For non-zero $r_0$ the
radial coordinate extends from $r=r_0$ (for $\phi=0$) to infinity
(where $\phi(r)$ approaches one of the asymptotic values
$\phi_\pm$). The induced metric has no $AdS_5 \times S^3$ structure
anymore. As was advertised in the Introduction the $D7$-branes do
not reside at the antipodal points on the $(\theta,\phi)$
two-sphere. This is not really surprising since there is a conic
singularity at the tip, so the  $S^2$ does not shrink smoothly. This
is in contrast to the low-temperature confining phase of the
Sakai-Sugimoto model, where the $x_4$ circle smoothly shrinks to
zero size resembling the cigar geometry. For a non-orbifolded
$\mathbb{R}^2$ plane spanned by the polar coordinates $(r,\phi)$ a
straight line is given by $\cos (\phi)=r_0/r$, where, again, $r_0$
is the minimal distance between the origin and the line. The
equation (\ref{Solution:eq}) has a similar form, where the $4$th
power and the $4/\sqrt{6}$ factor are both artifacts of the conic
singularity of the $6d$ conifold.

Before closing this section let us notice that (\ref{Solution:eq}) means that we
have a family of classical solutions with different parameter $r_0$, but with the
\emph{same} boundary values $\phi_+$ and $\phi_-$ at $r \to \infty$. This implies that
once we consider a perturbation theory around the classical profile we should find
a massless mode related to the variation of $y_\textrm{cl}$ with respect to $r_0$.
Since for $r_0>0$ the induced metric has no $AdS_5$ factor, the conformal
symmetry of the dual gauge theory should be broken in this case.
The massless mode, therefore, is just the Nambu-Goldstone boson of the broken conformal invariance.
In the next section we will see that this mode indeed appears in the perturbative expansion.

%%%%%%%%%%%%%%%%%%%%%%%%%%%%%%%%%%%%%%%%%%%%%%%%%%%%%%%%%%%%%

\section{\bf Spectrum of mesons}
\label{Spectrum}

In this section we will calculate the spectrum of mesons. We will begin with the scalar mesons
coming from the variations of the transversal coordinates and will end up with the vector mesons related to the expansion in term of the $D$-brane gauge fields. In both cases we will ignore
the non-trivial three-sphere modes.

We start with an observation that the ``polar" coordinates $r$ and
$\phi$ we used in the profile equation (\ref{Solution:eq}) do not
provide a convenient parameterization of the embedding. As we have
already seen, for a fixed value of $r$ the equation
(\ref{Solution:eq}) has two solution corresponding to the two
branches of the brane. We therefore cannot use $r$ as an independent
coordinate if we want to distinguish between the branches. Moreover,
at $r=r_0$ the derivative $\partial_r \phi(r)$ blows up making the
expansion around the classical configuration somewhat problematic.
On the other hand, using $\phi$ as an independent coordinate we find
that  the expansion becomes very complicated and the derivative
$\partial_\phi r(\phi)$ diverges now at $\phi=\phi_\pm$. To
summarise, we need a new set of coordinates which properly describes
the two branches of the $D7$-brane and also renders the profile
(\ref{Solution:eq}) in a non-singular form.

We found that the following ``Cartesian" coordinates do the job:
\beq
y = r^4 \cos \left( \frac{4}{\sqrt{6}} \phi\right)
\qquad \textrm{and} \qquad
z = r^4 \sin \left( \frac{4}{\sqrt{6}} \phi\right).
\eeq
With the malice of hindsight we have used the same notation as in the
original Sakai-Sugimoto paper \cite{Sakai:2004cn}.
Along the configuration (\ref{Solution:eq}) the coordinate
$y$ remains fixed $y_\textrm{cl}=r_0^4$, while $z$ takes all real values.
Furthermore, for positive and negative $z$
we have two different branches of the brane.
The situation thereof is a generalisation of the coordinates used in \cite{Sakai:2004cn},
where only the $y_\textrm{cl}=0$ case was studied.
From now on we will use $z$ together with the space-time coordinates $x_\mu$ and the Maurer-Cartan forms
$f_i$ to parameterize the world-volume of the $D7$-brane.
In particular, the induced $8d$ metric on the brane is:
\beq
\label{metric8dyz:eq}
\d s_{(8)}^2 = \frac{r^2}{R^2} \d x_\mu \d x^\mu
  + R^2 \left(
                \frac{ \left( z^2 + 2 r_0^8  \right)}{16 r^{16}}  \d z^2
                - \frac{\sqrt{6} r_0^4}{12 r^8} \d z f_1
                + \frac{1}{6} \left( f_1^2 + f_2^2 \right) + \frac{f_3^2}{9}
        \right),
\eeq
where $r=r(z)$ is given by:
\beq
\label{rrz:eq}
r^8 = z^2 + r_0^8.
\eeq
In the rest of the section we will use the coordinates $y$ and $z$ to compute
the scalar and the vector mesonic spectra.

\subsection{\bf Scalar mesons}

Plugging $y=y_\textrm{cl}+\delta y(x_\mu,z)$ and $\theta=\theta_\textrm{cl}+\delta\theta(x_\mu,z)$
into the DBI action (\ref{DBI:eq}), expanding around the
classical solution $(y_\textrm{cl},\theta_\textrm{cl})=(r_0^4,\frac{\pi}{2})$
and integrating over the three-sphere,
we arrive at the following action
for the fluctuation fields $\delta y(x_\mu,z)$ and $\delta\theta(x_\mu,z)$:
\bea
\label{dL:eq}
\delta S_\textrm{DBI} &=& -\frac{2 \pi^2}{72} \mu_7 \int \d x_\mu \d z
                    \Bigg\{
            \frac{1}{2}  \left( \partial_z \delta y\right)^2 +
            \frac{R^4}{32r^{10}} \left( \partial_\mu \delta y \right)^2 + \\
    &&     \qquad \qquad
          + \frac{4}{3} r^8 \left( \partial_z \delta \theta \right)^2 -
            \frac{r_0^8}{2 r^8}  \delta \theta^2 +
            \frac{R^4}{12 r^{2}} \left( \partial_\mu \delta \theta \right)^2
                    \Bigg\},
\nonumber
\eea
where $r$ is given by (\ref{rrz:eq}).

Let us start with the $\delta y(x_\nu,z)$ field.
As usual in a meson spectrum calculation we will assume
that $\partial_\mu \partial^\mu \delta y (x_\nu,z) = M^2 \, \delta y (x_\nu,z) $,
where $M$ is the $4d$ mass.
Introducing  a dimensionless variable $x$ and a parameter $\lambda$:
\beq
x=\frac{z}{r_0^4} \qquad  \textrm{and} \qquad \lambda=\frac{R^2 M}{r_0}
\eeq
we obtain the following Schr\"odinger-like equation:
\beq
\label{ddy:eq}
\partial_x^2  \,\,\delta y + \frac{\lambda^2}{16 (1+x^2)^{5/4}} \cdot \delta y= 0.
\eeq In order for  the expansion in terms of $\delta y$ to be
well-defined the function as well as its derivatives have to be
regular (non-divergent) for any value of $x$. The function should
also be normalisable at $x\to 0$ and $x \to \pm \infty$. This
immediately implies that $\lambda^2 \geqslant 0$ (and so $M^2
\geqslant 0$), since otherwise the potential in (\ref{ddy:eq}) is
everywhere positive and so there are no normalisable solutions.
Notice also that the potential in (\ref{ddy:eq}) is even under $x
\to -x$. Thus we expect to find pairs of even and one odd solutions.
Indeed, near $x=0$ we have $\delta y \sim 1 + \mathcal{O}(x^2)$ or
$\delta y \sim x + \mathcal{O}(x^3)$. On the other hand, for $x \to
\infty$ we find that $\delta y \sim 1$ or $ \delta y \sim x$.
Clearly we have to keep only the former option (the latter solution
is also non-normalisable for the action (\ref{dL:eq})).

Before applying a numerical method to solve (\ref{ddy:eq}) for $M>0$
we would like to point out that the equation is easily solvable for
$M=0$. The solutions are $\delta y =1$ and $\delta y =x$. The linear
solution is non-normalisable, so we are left only with the first
option. This constant solution is exactly the Nambu-Goldstone boson
of the broken conformal symmetry we have predicted in the end of the
previous section. Consistently this massless mode is $r$-independent
since, as was  already discussed above, it comes from the
$r_0$-derivative of the classical configuration $y_\textrm{cl}$,
which in turn is $r$-independent.

We now want to solve (\ref{dL:eq}) with $M>0$ for the entire range
of $x$ by gluing one of the two solutions at $x=0$ with the
non-divergent solution at infinity. This is, of course, possible
only for discrete values of $\lambda_\n$, which we found by means of
the ``shooting technique". Setting the even ($\delta y(0)=1$,
$\delta y^\pr (0)=0$) or the odd ($\delta y(0)=0$, $\delta y^\pr
(0)=1$) boundary conditions at $x=0$, we solved the equation
numerically fixing $\lambda$ by allowing only the normalisable
(finite $\delta y$) solution for $x \gg 1$. As we have already
argued the even (odd) initial conditions at $x=0$ lead to even (odd)
solutions of (\ref{ddy:eq}) and vice versa.

We found:
\beq
\lambda_\textrm{n}^{PC} = 4.03^{--},  \, 5.55^{++} , \,
                     7.01^{--},  \, 8.43^{++} , \,
                     9.83^{--}, \, 11.21^{++} \, \ldots
\eeq Before explaining the  parity ($P$) and charge conjugation ($C$)
  assignments let us analyse the $\delta \theta(x_\nu,z)$ field.
The same procedure as for $\delta y$ leads to: \beq
\label{ddtheta:eq}
\partial_x \left( \left(1+x^2 \right) \partial_x \delta \theta \right)
   + \frac{1}{8} \left( \frac{3}{1+x^2}  + \frac{\lambda^2}{2 (1+x^2)^{1/4}}
                                                              \right)\delta \theta = 0.
\eeq
At infinity we have $\delta \theta \sim 1/x$ or $\delta \theta \sim 1$, only the former
of which is acceptable, while the latter is now non-normalisable
(see the last term in (\ref{dL:eq})).
Near $x=0$ we have $\delta \theta \sim 1$ or $ \delta \theta \sim x$
exactly like for the $\delta y$ field. Again,
both solutions are convergent and give rise to even and odd solutions respectively.
For this field the spectrum is:
\beq
\lambda_\textrm{n}^{PC} = 2.61^{-+},  \, 4.39^{+-}  ,\,
                     5.81^{-+},  \, 8.63^{+-} ,\,
                     8.63^{-+},  \, 11.38^{+-} \, \ldots
\eeq

We can now compare these scalar meson spectra to the corresponding
spectra of \cite{Sakai:2004cn} and
\cite{Mintkevich:2008mm}. We observe that in the latter two models there are
scalar states with $0^{++}$ and $0^{--}$ whereas in our model there
are states with all the four combinations of $P$ and $C$. In all
models there are $0^{--}$ low lying meson states that do not occur in
nature.

Let us now explain the parity and the charge conjugation properties of the modes.
Our analysis will be very similar to  \cite{Sakai:2004cn}.
We can fix the $4d$ parities by requiring the $8d$ action on the $D7$ branes
to be $C$ and $P$ invariant. After KK reduction on $S^3$ the $5d$
$P$-parity transformation  reads $(x_i,z) \to (-x_i,-z)$, while the
charge conjugation implies both $z \to -z$ and $A \to -A$
(or $A \to -A^\textrm{T}$ in the non-Abelian case, see \cite{Sakai:2004cn}).
Since all the fields appear quadratically in the DBI part we will
not be able to determine the parities from this part of the action.
There is a non-trivial RR $4$-form potential $C_4$ in the background, however, and so
we have also two Chern-Simons (CS) terms in the action. Both terms do not modify the spectrum calculation,
since in the Abelian case they are at least cubic in the field fluctuations, but nevertheless these terms
reveal the parity and the charge conjugation transformations of the fields. The first term is:
\beq
\int F \wedge F \wedge C_4, \qquad \textrm{with} \qquad
    C_4 \sim r^4 \d x_0 \wedge \d x_1 \wedge\d x_2 \wedge\d x_3.
\eeq
Here $F$ is the gauge field strength on the brane\footnote
{To be precise in the Abelian case $F \wedge F$ is a total derivative and so the term does not
modify the equations of motion.
In the non-Abelian case we will have to replace  $F \wedge F$
by $\textrm{Tr} (F \wedge F)$.}.
This term does not provide any new insight, since it has no $\delta \theta$
or $\delta y$ dependence. The second CS term is due to the Hodge dual of $C_4$,
which by definition satisfies $\d \widetilde{C}_4 = \star_{10} \d C_4$.
Up to a gauge transformation we have:
\beq
 \widetilde{C}_4 \sim \cos \theta \, \d \phi \wedge \omega_1^\pr \wedge \omega_2\pr  \wedge \omega_3\pr,
\eeq where $w_i^\pr$ are the $SU(2)$ Maurer-Cartan forms we have
introduced in Section \ref{Theconfiguration}. This CS term yields
the following coupling in the $5d$ action: \beq \int F \wedge F
\wedge \delta \theta(y_\textrm{cl} + \delta y - z \partial_z \delta
y )  \d z, \eeq where we kept only the two lowest terms in the
perturbative expansion. We see that $\delta y$ should transform
exactly like $y_\textrm{cl}$, which is constant and so clearly both
charge conjugation  and parity are even. On the other hand , $\delta
\theta$ is $C$ even and $P$ odd. The $4d$ parities of the $\delta y$
and $\delta \theta$ modes depend  on the solution choice in
(\ref{ddy:eq}) and (\ref{ddtheta:eq}). For example, for even
solutions of (\ref{ddtheta:eq}) we get $0^{-+}$ modes, while odd
solutions correspond to $0^{+-}$ modes.

\subsection{\bf Vector mesons}

Since in this paper we consider only a single probe brane, the first
non-trivial contribution in the $F$-expansion of the DBI action
yields only the standard $F \wedge \star F$ Abelian term. There is
also an $F \wedge F$ term coming from the $C_4$ part of the
Chern-Simons action, but  this term is a total derivative that does
not modify the equations of motion. Because we are interested only
in the three-sphere independent modes we will ignore gauge fields
with legs along the $S^3$ and will assume also that the remaining
fields depend only on the coordinates $z$ and $x_\nu$. The action
then reduces to a $5d$ Maxwell action with a $5d$ background metric,
which we can find from (\ref{metric8dyz:eq}) ignoring the $S^3$
directions. The action is: \beq \label{Maxwell:eq} S = -T^\prime
\int \d x^4 \d z \big( C(z) F_{\mu \nu}  F^{\mu \nu} + 2 D(z) F_{\mu
z} F^\mu_z \big), \eeq where we absorbed various  numerical and
dimensionful constants in $T^\prime$, the space-time indices $\mu$,
$\nu$ are contracted with the Minkowskian metric and: \beq
\label{CD:eq} C(z) = \frac{R^4}{(z^2+r_0^8)^{1/2}} \propto
\sqrt{-g_8} \left(g_8^{\mu \nu}\right)^2 \,\,\, \textrm{and}
\,\,\,\, D(z) = 16 (z^2+r_0^8)^{3/4} \propto \sqrt{-g_8} g_8^{\mu
\nu} g_8^{zz}. \eeq Here $g_8$ stands for the $8d$ metric
(\ref{metric8dyz:eq}). Next we consider the following mode
decomposition of the fields: 
\beq 
A_\mu (x,z) = \sum_{\n}
a^\n_\mu(x) \alpha^\n(z) \quad \textrm{and} \quad
 A_z (x,z) = \sum_{\n} b^\n (x) \beta^\n(z).
\eeq
With this decomposition the field strength reads:
\beq
F_{\mu \nu} = \sum_{\n} f^\n_{\mu\nu} (x) \alpha^\n(z)  
\quad \textrm{and} \quad
F_{\mu z} = \sum_{\n}\left( \partial_\mu b^\n (x) \beta^\n(z) - a^\n_\mu(x) \partial_z \alpha^\n(z) \right),
\eeq
where $f_{\mu\nu}=\partial_\mu a_{\nu}-\partial_\nu a_{\mu}$. Substituting this back into the action
(\ref{Maxwell:eq}) we receive:
\bea
\label{MaxwellModes:eq}
S &=& - T^\prime \int \d x^4 \d z \sum_{\m,\n} \Big( C(z) f_{\mu \nu}^\n {f^\n}^{\mu \nu} \alpha^\n \alpha^\m   + \\
  & & \quad + 2 D(z) \left( \partial_\mu b^\n \partial^\mu b^\m \beta^\n \beta^\m
                                        + a^\n_\mu a^{\m\mu} \partial_z \alpha^\n \partial_z \alpha^\m
   -2 \partial_\mu b^\n a^{\m\mu} \beta^\n \partial_z \alpha^\m    \right) \Big).  \nonumber
\eea
Following \cite{Sakai:2004cn} we first consider the equation of motion and the normalization condition for $\alpha^\n(z)$:
\beq
\label{alpha:eq}
-\frac{1}{C(z)} \partial_z \left( D(z) \partial_z \alpha^\n (z) \right) = M_\n^2 \alpha^\n(z)
\,\,\, \textrm{and} \,\,\,\,
T^\prime \int_{-\infty}^\infty \d z C(z) \alpha^\n (z) \alpha^\m (z) = \delta_{\n\m}.
\eeq
Here the equation of motion is derived from the third term in (\ref{MaxwellModes:eq}),
while the normalization is dictated by the first term. Using both equations in (\ref{MaxwellModes:eq})
we get rid of the $z$-dependence of the first and the third terms obtaining
this way the standard  $4d$ kinetic and mass term for the
gauge fields $a^\n_\mu$'s. We have meanwhile ignored the $\beta^\nu(z)$ modes. The reason for that is the absence of
a kinetic term for these modes. This means that we only have to impose a right normalization for $\beta^\n(z)$'s. Remarkably,
the following simple substitution:
\beq
\beta^\n(z) = \frac{\partial_z \alpha^\n(z)}{M_\n}
\eeq
does the job. With the help of (\ref{alpha:eq}) the second term in (\ref{MaxwellModes:eq}) provides a
standard kinetic term $\partial_\mu b^\n \partial^\mu b^\n $ for the scalar fields $b^\n$'s, while the last term in
(\ref{MaxwellModes:eq}) reduces to the form $-2\partial_\mu b^\n {a^\n}^\mu$.
It turns out that both terms can be eliminated by the gauge transformation:
\beq
a_\mu^\n \longrightarrow  a^\n_\mu +\frac{\partial_\mu b^\n}{M_\n}.
\eeq
This seems to complete the analysis, meaning that there are no scalars in the final $4d$ action, only the gauge fields
$a^\n_\mu$. Yet there is a trap here: we just overlooked an additional normalisable
mode $\beta^\mathbf{0}(z)$, which is orthogonal to all other modes
${\beta^\n (z) \propto \partial_z \alpha^\n(z)}$ for all $\n \geqslant 1$ with respect to the scalar product
defined by the second term in (\ref{MaxwellModes:eq}). This mode is $\beta^\mathbf{0}(z)=\kappa/D(z)$.
We can easily check that:
\beq
\int_{-\infty}^\infty \d z D(z) \beta^\mathbf{0}(z) \beta^\n(z) = \frac{\kappa}{M_\n}
                          \int_{-\infty}^\infty \d z \partial_z \alpha^\n(z)=0.
\eeq
The constant $\kappa$ has to be fixed by the normalization of the mode $\beta^\mathbf{0}(z)$:
\beq
\label{normalization:eq}
\frac{1}{\kappa^2} = 4 T^\prime \int_{-\infty}^\infty \frac{\d z}{D(z)}.
\eeq
Plugging $\beta^\mathbf{0}(z)$ into the action we find an additional scalar kinetic term
$\partial^\mu b^\mathbf{0} \partial_\mu b^\mathbf{0}$ that cannot be eliminated by any gauge transformation.
To summarise, we find that the $4d$ action consists of the massive gauge fields $\a_\mu^\n$ and the massless
scalar  $b^{\mathbf{0}}$:
\beq
S_{4d} = -\int \d x^4 \left( \half \partial^\mu b^\mathbf{0} \partial_\mu b^\mathbf{0} +
          \sum_{\n \geqslant 1}  \left(
          \frac{1}{4} f_\n^{\mu\nu} f^{\n\mu\nu} + \half M_\n^2 a_\mu^\n a^{\n\mu} \right)  \right).
\eeq Following the discussion in Introduction we will identify
$b^\mathbf{0}$ as the Goldstone boson of the broken chiral symmetry.
This implies that the we should anticipate this mode only for $r_0 >0$, 
namely for the $U$-shape of two smoothly merging $D7$-branes,
but not for $r_0=0$ which corresponds to the $V$-shape of two
separate branes. The answer to this puzzle is encoded in the
convergence of the integral in (\ref{normalization:eq}). For $r_0=0$
we have $D(z)=16 \cdot z^{-3/2}$ and the integral
(\ref{normalization:eq}) diverges at $z=0$, so, as predicted, the
massless mode does not exist for the $V$-shape. On the other hand,
the integral is finite for $r_0 > 0$ as expected.

Our last goal in this section is to find the spectrum of the massive vector mesons. To this end we have to solve
the first equation in (\ref{alpha:eq}). Proceeding the same way like with the scalar mesons we obtain the following results:
\beq
\lambda_\textrm{n} = 2.03^{++},\quad 3.32^{--}, \quad  4.71^{++},
               \quad 6.05^{--}, \quad  7.41^{++},  \quad 8.76^{--},  \ldots
\eeq
Here the the parity and the charge conjugation properties are identified exactly like in the
Sakai-Sugimoto model \cite{Sakai:2004cn}. In particular, the massless mode $b^\mathbf{0}$
is $0^{-+}$.

%%%%%%%%%%%%%%%%%%%%%%%%%%%%%%%%%%%%%%%%%%%%%%%%%%%%%%%%%%%%%

\section{The dual gauge theory}
\label{GT}

In this section we will analyse the dual gauge theory.
As we have already mentioned in Section \ref{Theconfiguration}
the K\"ahler quotient coordinates $\mathfrak{z}_i$ of the conifold
correspond to the chiral bi-fundamentals $A_i$ and $B_i$ in the quiver gauge theory.
To be more specific, we have $u \propto \left( A_1,A_2 \right)^\textrm{T}$
and $v^\star \propto \left( B_1,B_2 \right)^\textrm{T}$, see (\ref{zzAB:eq}).
In this paper we used $X \in SU(2)$ and $v$ to parameterize
the three- and the two-spheres of the conifold and our embedding
looks like two separate points on $S^2$. The position of these points depends
on the radial coordinate for $r_0 >0$
(broken conformal and chiral symmetries)
and is fixed for $r_0=0$ (un-broken symmetries).

For the embedding to be supersymmetric 
(namely to preserve four out of the eight supercharges of the background)
it has to be given by
a holomorphic function \cite{Becker:1995kb} (see also \cite{Arean:2004mm}).
It is easy to check that for $r_0 >0$ the embedding is explicitly non-holomorphic.
Let us now address the $r_0=0$ case.
Since the conifold inherits the complex structure of $\IC^4$ we conclude that the $r_0=0$
embedding is supersymmetric if and only if one has $B_1=0$ or $B_2=0$ along the brane
(which is the same as
$\mathfrak{z}_3=0$ or $\mathfrak{z}_4=0$). This, however, describes two antipodal
points on the 2-sphere parameterized by $v$ and we have demonstrated that there is no such
solution\footnote{\label{AP}
Recall that $B_1=0$ and $B_2=0$ correspond to $v=(0,1)^\textrm{T}$ and 
$v_2=(1,0)^\textrm{T}$ respectively. These points are the north and the south poles 
of the $2$-sphere described by $v$.}. 
Instead we found that the angle difference is $\frac{\sqrt{6}}{4} \pi$.
To conclude, the embedding breaks supersymmetry for any $r_0$.

\begin{figure}[!h]
\begin{center}
\includegraphics[scale=0.6
]{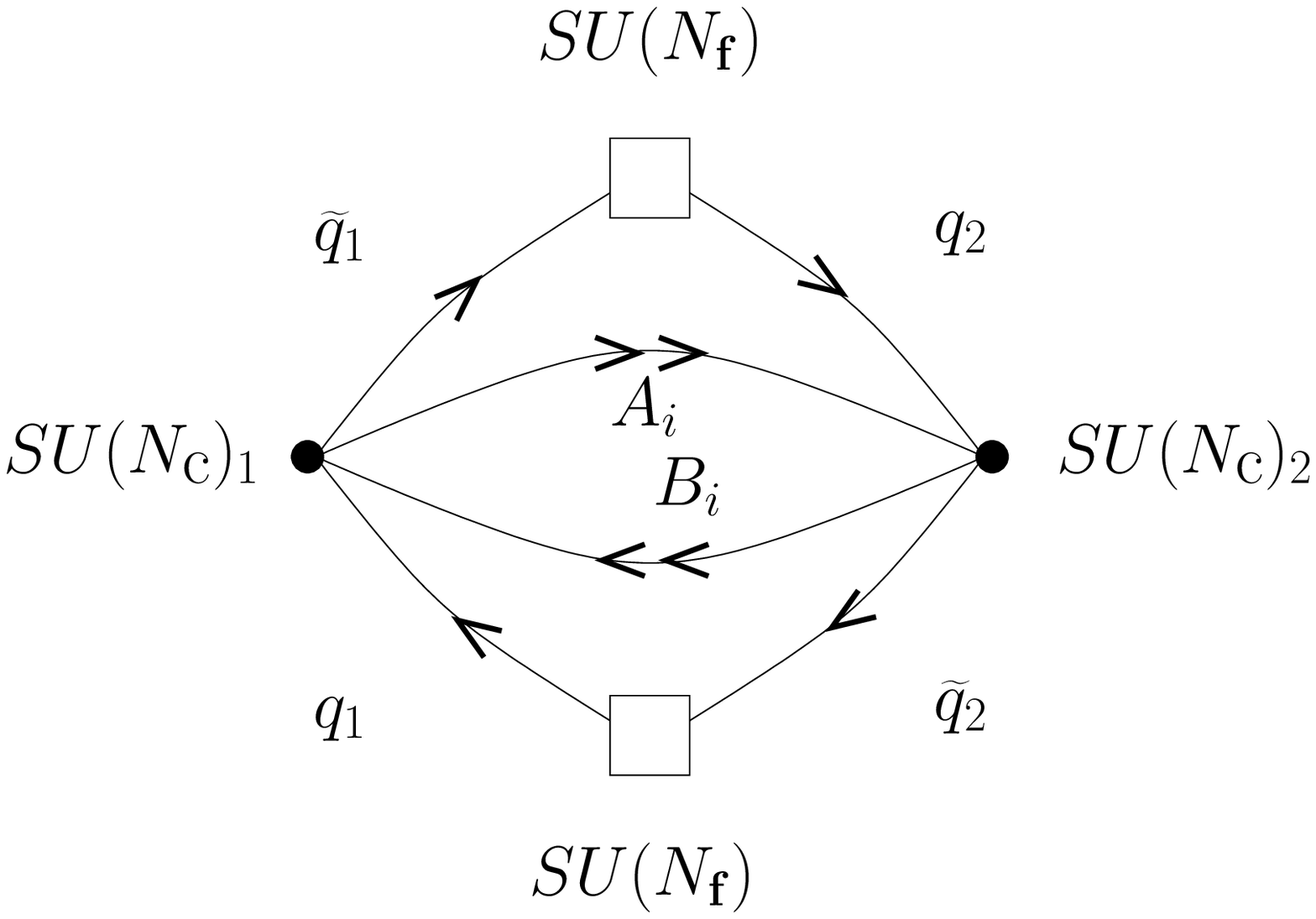}
\caption{The quiver diagram of the supersymmetric embedding elaborated in \cite{Ouyang}.
Here dots denote the gauge groups and the boxes correspond to the global flavour symmetries.
Notice that there is no anomaly, as for each node the number of incoming and outgoing arrows
are equal.}
\label{QuiverOuyangLike}
\end{center}
\end{figure}

The fact that there is no supersymmetric antipodal configuration
matches, to some extent, the quiver gauge theory expectations. To
see this, let us first consider the holomorphic embedding studied in
\cite{Ouyang}. In terms of the bi-fundamentals it is given by: 
\beq
\label{Ouyang:eq} 
A_1 B_1 = \mu 
\eeq 
and we will put $\mu=0$ for
simplicity. In this case it is straightforward to find the quiver
diagram and the flavour part of the superpotential. The quiver of
Figure \ref{QuiverOuyangLike} and the additional part in the
superpotential is \cite{Ouyang}: 
\beq 
\label{OuyangW:eq}
\Delta W = q_2 B_1 \widetilde{q}_1 + q_1 A_1 \widetilde{q}_2. 
\eeq 
Higgsing the fields $A_1$ and $B_1$ one
finds massive quarks, while the requirements for the quarks to be
massless leads to the $A_1B_1=0$ embedding (see \cite{Me,Ouyang}).

Notice now that the same approach will \emph{not} work for the $B_1 B_2=0$ embedding,
which describes $D7$ and anti-$D7$ at the antipodal points on the two-sphere (see Footnote
\ref{AP}).
This is because in order to simultaneously include the terms ${q_2} B_1 \widetilde{q}_1$
and $\widetilde{q}_2 B_2 q_1$ in the superpotential we will have to
invert the arrows of $\widetilde{q}_2$ and $q_1$ in the diagram on Figure \ref{QuiverOuyangLike}.
This, however, will produce an anomalous quiver diagram,
since the number of incoming and outgoing arrows (for either node $1$ or $2$) will be different.
We see that as expected we cannot add flavours to the gauge theory in a way that will correspond
to the antipodal brane configuration.

\begin{figure}[!h]
\begin{center}
\includegraphics[scale=0.7
]{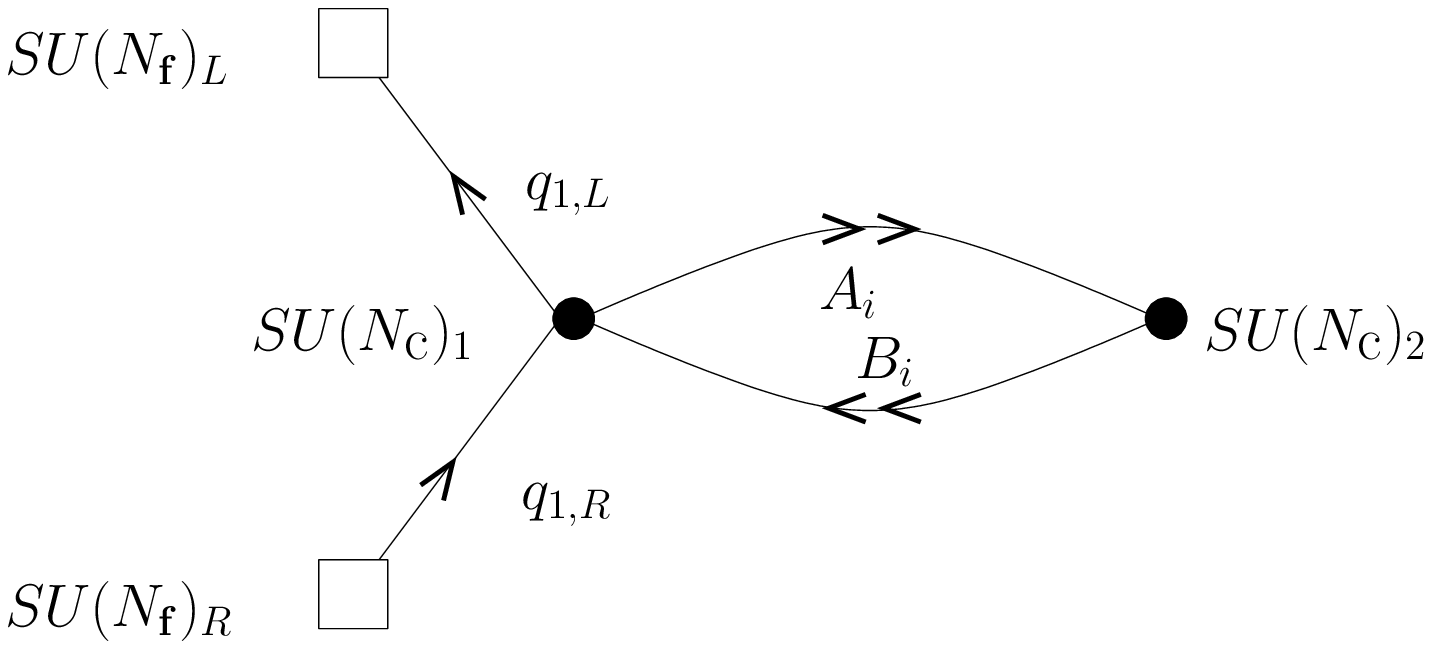}
\caption{A quiver diagram that doesn't respect the $\IZ_2$ symmetry.}
\label{QuiverOneNode}
\end{center}
\end{figure}

We argued in Section \ref{Theconfiguration} that our $D7$-brane configuration breaks the $\IZ_2$
symmetry. Recall that this symmetry interchanges the gauge groups and so the quiver
diagram on Figure \ref{QuiverOuyangLike}
is obviously $\IZ_2$-invariant.
This is in agreement with the definition of the embedding (\ref{Ouyang:eq}), which is invariant under
$A_i \leftrightarrow B_i$. So we may wonder whether this is the right diagram for our embedding.
For instance, we can consider a different quiver diagram presented on Figure \ref{QuiverOneNode},
where the quarks interact only with one of the two gauge groups.
Although this diagram breaks the $\IZ_2$ and seems to be a perfect candidate for our
model, it does not allow actually for \emph{any} interaction between the quarks and the bi-fundamentals.
Indeed, there are only two possible interactions consistent with the quiver diagram.
A term like\footnote{Since our setup is non-supersymmetric we write terms in the \emph{potential} and not in
the superpotential, still using the same notations for the regular (bosonic and fermionic) fields as for the
superfields.}
$q_{1_L} \Phi \bar{q}_{1_L}$, where $\Phi$ is an adjoint field of the form
$\Phi=A_i B_j$, is not Lorentz invariant, while a term  $q_{1_L} \Phi q_{1_R}$ breaks chiral symmetry \emph{explicitly}.

An additional possibility we may consider is the quiver diagram of \cite{Me,Benini:2007gx}. In this case,
the quarks and the anti-quarks of the same $SU(N_\textrm{f})$ couple to the same gauge group. Clearly this
is not the right diagram, since for any chiral symmetry breaking setup we
need left and right quarks with the same gauge group but with different flavour groups $SU(N_\textrm{f})_L$ and
$SU(N_\textrm{f})_R$.

We propose therefore that Figure
\ref{QuiverOuyangLike} is the quiver diagram corresponding to our embedding
although it does not break the $\IZ_2$ invariance. 
Of course, for our non-supersymmetric model the arrows 
on the diagram are not related anymore to chiral superfields,
but rather to fermions (for $q$'s) and bosons (for $A_i$'s and $B_i$'s).
We suggest that the $\IZ_2$ breaking will come from the explicit terms
in the potential, which unfortunately we were not able to find.

One may raise the question whether our model really describes chiral symmetry breaking,
namely do we have Weyl or Dirac spinors for each one of the $D7$-branes.
The chiral symmetry breaking scenario can be realized only for the former case.
Let us demonstrate that this is indeed what we have. For $\mu=0$ the embedding (\ref{Ouyang:eq})
introduced in \cite{Ouyang}
describes two branches $A_1=0$ and $B_1=0$. Each branch describes an $S^3$ on $T^{1,1}$. Unlike in our setup, these
three-spheres intersect along an $S^1$ on the base of the conifold. Indeed,
plugging $A_1=B_1=0$ into the  $D$-term condition 
$\vert A_1 \vert^2+\vert A_2 \vert^2-\vert B_1 \vert^2-\vert B_2 \vert^2=0$
we find that $\vert A_2 \vert=\vert B_2 \vert$. Recall that we also have to quotient $A_2$ and $B_2$
by the $U(1)_K$, and so the intersection of $A_1=0$ and $B_1=0$ is a $2d$ cone parameterised by 
the gauge invariant combination $A_2 B_2$, which in turn means that on the $5d$ base $T^{1,1}$
the intersection looks like $S^1$. This is in contrast to our model where the 
two branches look like two \emph{non-intersecting} $S^3$ with opposite orientations
(we believe that for (\ref{Ouyang:eq}) the orientations of the spheres are the same
since the embedding is supersymmetric).  Still, we can consider only 
the $B_1=0$ branch of this holomorphic embedding.  This branch looks exactly 
like one of the branes in our model. This brane alone is supersymmetric
and we can assume that its contribution to the superpotential is just the first term in 
(\ref{OuyangW:eq}).  The chiral multiplets $\widetilde{q}_1$ and $q_2$, however,
both have \emph{left} Weyl fermions. The other branch of our configuration
is an anti $D7$-brane, since it 
has an opposite orientation and breaks supersymmetry. Thus 
it should have \emph{right} Weyl fermions instead. 
The contribution of these fermions to the \emph{potential}
should be similar to the potential term one can derive from the first term in (\ref{OuyangW:eq}).
Instead of $B_1$ this term should include the field 
$\cos(\alpha) \bar{B_1} + \sin(\alpha) \bar{B_2}$,
where $\alpha=\frac{\sqrt{6}}{2}\pi$ is the angle between the two points on the $2$-sphere
corresponding to the brane and the anti-brane. 
The contribution to the potential of the $D7$-brane and the anti $D7$-brane
will preserve different supersymmetries and so the entire setup will be non-supersymmetric.
To summarise, our brane and anti-brane have left and right fermions respectively
and so the merging of the branes indeed corresponds to chiral symmetry breaking.
It will be very intersting to calculate the potential of our model following
the arguments above.

%%%%%%%%%%%%%%%%%%%%%%%%%%%%%%%%%%%%%%%%%%%%%%%%%%%%%%%%%%%%%

\section{A model with no chiral symmetry breaking}
\label{z4}

In this section we will examine a different embedding originally proposed in \cite{Me} for the
deformed conifold. We focus on this embedding merely because similarly to our model it preserves
one $SU(2)$ factor of the isometry group making the analysis much simpler. We believe that on the same footing
we could have studied an alternative embedding like, for example, the one considered in \cite{Ouyang}
still arriving at the same conclusions.

We would like to demonstrate that the embedding of \cite{Me} does not look like a $U$-shape configuration
that smoothly merges into a single brane, which for a specific value of the embedding parameter splits into a pair
of two non-intersecting branes. In other words this model does not possess any chiral symmetry breaking.
We will then argue that the vector meson spectrum in this case has no massless
Goldstone boson in accordance with the expectations.

The spectrum of the vector mesons has already been calculated in \cite{Me} for the deformed conifold
(the Klebanov-Strassler model \cite{Klebanov:2000hb})
and no massless modes have been found there. Here we want to repeat the computation for the singular conifold
(the Klebanov-Witten model \cite{Klebanov:1998hh})
following the steps presented in the Section \ref{Spectrum}.

The embedding we are interested in is:
\beq
\label{z4mu:eq}
z_4=\mu, \qquad \textrm{where} \qquad W = \left( \begin{array}{cc} z_3 + i z_4 & z_1 - i z_2 \\
                                                z_1 + i z_2 & -z_3 + i z_4 \end{array} \right)
\eeq
is the matrix we used to define the conifold geometry.
Since $2 i z_4 = \textrm{Tr} W$ the profile (\ref{z4mu:eq}) preserves the diagonal $SU(2)_D$ isometry
that acts like $W \to S_D W S_D^\dagger$.

In the $z_i$ coordinates the conifold definition $\det W=0$ reads:
\beq
\label{Sumz2:eq}
\sum_{i=1}^4 z_i^2=0.
\eeq
From (\ref{Sumz2:eq}) we can understand the topology of the embedding. Let us first consider the $\mu=0$
case. Substituting $z_4=0$ into (\ref{Sumz2:eq}) and defining $u=z_1+iz_2$, $v=z_1-iz_2$ and $z=iz_3$ we obtain:
\beq
\label{Lens:eq}
uv-z^2=0,
\eeq
which is a definition of a $4d$ cone over the Lens space $L(2;1)=S^3/\IZ_2$.
We can arrive at the same conclusion using the results of Section \ref{Theconfiguration}.
We see that for $\mu=0$ the matrix $W$ is traceless and so (\ref{Wuv:eq}) implies that $v^\dagger u=0$
and so up to the gauge transformation (\ref{uvuv:eq}) we have
$u= \epsilon v^\star$ and so $W=\rho \epsilon v^\star v^\dagger$.
The $U(1)_K$ gauge symmetry (\ref{uvuv:eq}) is not broken completely, because $W$ is still invariant
under $v \to -v$. Recall that with no $U(1)_K$ quotient $v$ defines an $S^3$, so the
result of the $\IZ_2$ orbifold is the aforementioned Lens space $S^3/\IZ_2$.
Next, for $\mu \neq 0$ the zero in (\ref{Lens:eq}) is replaced by $-\mu^2$. This corresponds
to the deformation of the $\IZ_2$ singularity\footnote{Actually since the space defined by (\ref{Lens:eq})
is hyper-K\"ahler there is no way to distinguish between deformation and resolution.}.
The Lens space $L(2;1)=S^3/\IZ_2$ is an $S^1$ fibration over $S^2$ with Chern class $2$.
For $\mu=0$ the Lens space shrinks to zero at the tip, but for non-zero $\mu$ only the $S^1$ fiber
shrinks, while the $S^2$ approaches a finite size controlled by $\mu$. The shrinking of the $S^1$ cycle
occurs when the radial coordinate $\rho$ of the conifold reaches its minimal value $\rho_\textrm{min}=2 \mu$
along the brane.

To summarise, we saw that for $\mu=0$ at fixed radial coordinate the embedding looks like the Lens space
$S^3/\IZ_2$ and for $\mu \neq 0$ the $U(1)$ fiber of the Lens space shrinks at $\rho=\rho_\textrm{min}$,
where the embedding looks like $S^2$.
Clearly the situation here does not resemble our setup. There are no separate branches of the $D7$-brane for $\mu=0$
that merge into a single configuration if we put $\mu \neq 0$.

In order to analyse the vector meson spectrum we will need the $8d$ induced metric
of the embedding (\ref{z4mu:eq}). For the deformed conifold
this metric was found in \cite{Me}. To get the induced metric for the un-deformed conifold we only have
to take the $\varepsilon \to 0$ limit, where $\varepsilon$ is the conifold deformation parameter.
The calculation is quite simple and here we report only the final result, referring the reader to \cite{Me}
for further details. The induced metric is:
\begin{eqnarray}
\d s_{(8)}^2 &=& \frac{r^2}{R^2} \d x_\mu \d x^\mu + \frac{R^2}{r^2} \d r^2 +
          R^2 \Bigg( \frac{1}{6} \bigg( h_1^2 +h_2^2 +  (h_1 - \partial_r  \gamma \d r)^2 +   \nonumber \\
 &&  \qquad  +(h_3 \sin \gamma + h_2 \cos \gamma)^2  \bigg) + \frac{1}{9} \big(h_3(1+ \cos \gamma) - h_2 \sin \gamma \big)^2
     \Bigg).
\end{eqnarray}
Here $h_i$ are the $SU(2)_D$ Maurer-Cartan forms and $\gamma=\gamma(r)$ satisfies:
\beq
\label{gamma:eq}
\sin \left( \frac{\gamma(r)}{2} \right) = \left( \frac{r_\textrm{min}}{r} \right)^{3/2}
\quad \textrm{with} \quad r_\textrm{min}=\frac{3^{1/2}}{2^{1/6}} \mu^{2/3},
\eeq
where $r_\textrm{min}$ is the minimal value of $r$ along the brane.  
In particular, it follows from (\ref{gamma:eq}) that for $\mu=0$
we get $r_\textrm{min}=0$ and $\gamma(r)=0$ for any $r$. In this case the metric is identical
to the metric in (\ref{metric6:eq}) for $r_0=0$ and describes $AdS_5 \times ``S^3"$ (see Footnote 
\ref{squashed}).

We are now in a position to analyse the integral (\ref{normalization:eq}) for the embedding $z_4 = \mu$.
As was explained in details in Section \ref{Spectrum} the massless mode exists only if the integral
in (\ref{normalization:eq}) converges.
Similar to (\ref{CD:eq}) we have:
\begin{displaymath}
\widetilde{D}(r) = \frac{r^3}{18} \left(  \frac{\cos^2 \gamma + 8 \cos \gamma +7}{1+\frac{1}{12} r^2 (\partial_r \gamma)^2}
    \right)^{1/2}
 \propto \sqrt{-g_8} g_8^{\mu \nu} g_8^{rr} .
\end{displaymath}
If $\mu=0$ then $\gamma(r)=0$ and $\widetilde{D}(r)=\frac{2}{9} r^3$. The integral (\ref{normalization:eq})
diverges and there is no massless vector meson exactly like in the $r_0=0$ case in our model.
The integral, however, diverges also for non-zero $\mu$.
To see this we have to find $\widetilde{D}(r)$ for $r \approx r_\textrm{min}$.
At this point $\gamma(r_\textrm{min})=\pi$.
Defining $\delta \gamma = \gamma - \pi$ and $\delta r = r - r_\textrm{min}$ 
we find from (\ref{gamma:eq}) that:
\beq
\delta \gamma \approx 2 \sqrt{3} \left( \frac{\delta r}{r_\textrm{min}} \right)^{1/2}.
\eeq
But then:
\beq
\widetilde{D}(r) \approx \frac{2}{3} \, r_\textrm{min}^2 \cdot \delta r
\eeq
and the integral (\ref{normalization:eq}) diverges logarithmically.
We therefore conclude that there is no massless vector meson in the $z_4=\mu$ setup and
so there is no chiral symmetry breaking in this case.

\section*{Acknowledgements}

It is a pleasure to thank Ofer  Aharony for very useful
conversations and for his comments on the manuscript. 
We are also grateful to Anatoly Dymarsky, 
Amit Giveon, Riccardo Argurio, Cyril Closset, Emiliano Imeroni, 
Francesco Bigazzi, Carlo Maccaferri, 
Chethan Krishnan, Jarah Evslin and especially Daniel Persson
for fruitful discussions.
The work of J.S was supported in part by a centre of excellence supported by the
Israel Science Foundation (grant number 1468/06), by a grant (DIP
H52) of the German Israel Project Cooperation, by a BSF grant,
by the European Network MRTN-CT-2004-512194 and
by European Union Excellence Grant MEXT-CT-2003-509661.

%\begin{appendix}
%\section*{\bf Appendix}
%\addcontentsline{toc}{section}{Appendix}

%\subsection*{{\bf A} \ }

\renewcommand{\theequation}{A.\arabic{equation}}

\newpage
%%%%%%%%%%%%%%%%%%%%%%%%%%%%%%%

\end{document}